\documentstyle[12pt]{article}
\def\be{\begin{equation}}
\def\ee{\end{equation}}
\def\ba{\begin{array}}
\def\ea{\end{array}}

\begin{document}
\title{ Chiral Quark Model with Configuration Mixing.}
\author{Harleen Dahiya  and Manmohan Gupta    \\
{\it Department of Physics,} \\
{\it Centre of Advanced Study in Physics,} \\
{\it Panjab University,Chandigarh-160 014, India.}}   
 \maketitle

\begin{abstract}
The implications of one gluon exchange generated configuration
mixing in the Chiral Quark Model ($\chi$QM$_{gcm}$) with SU(3) and
axial U(1) symmetry breakings are discussed in the context of proton
flavor and spin structure  as well as the hyperon $\beta$-decay parameters. 
We find that $\chi$QM$_{gcm}$ with SU(3) symmetry breaking is able to give a 
satisfactory unified fit for spin and quark distribution functions, with
the symmetry breaking parameters $\alpha=.4$, $\beta=.7$ and the mixing angle
$\phi=20^o$, both for  NMC and the most recent E866 data. In particular,
the agreement with data, in the case of
$G_A/G_V, ~\Delta_8$, F, D, $f_s$ and $f_3/f_8$, is quite striking.
\end{abstract}

It is well known that the chiral quark model ($\chi$QM)
{\cite{{manohar},{wein},{cheng}}}
with SU(3) symmetry is not only able to give a fair explanation of
``proton spin crisis'' {\cite{EMC}} but is also able to account for
the $\bar u-\bar d$ asymmetry {\cite{{NMC},{E866},{GSR1}}} as well
as the existence of significant strange
quark content $\bar s$ in the nucleon when the asymmetric
octet singlet couplings are taken into account  {\cite{st q}}.
Further, $\chi$QM with SU(3) symmetry is
also able to provide fairly satisfactory explanation for various
quark flavor contributions to the proton spin {\cite{eichten}}, baryon
magnetic moments {\cite{{cheng},{eichten}}} as well as the absence of
 polarizations of the
antiquark sea  in the nucleon {\cite{{song},{antiquark}}} .
However, in the case of hyperon decay parameters
 the predictions of the
$\chi$QM are not in tune with the data   {\cite{decays}}, for example,
in comparison to the experimental numbers   .21 and 2.17
the $\chi$QM with SU(3) symmetry predicts   $f_3/f_8$ and
$\Delta_3/\Delta_8$ to be $\frac{1}{3}$ and $\frac{5}{3}$ respectively.
It has been shown {\cite{{song},{cheng1}}} that when SU(3) breaking
 effects are taken into consideration within $\chi$QM, the
 predictions of the $\chi$QM regarding the above mentioned ratios
 have much better overlap with the data.

It is well known that constituent quark model (CQM) with one gluon 
mediated configuration mixing gives a fairly satisfactory explanation of
host of low energy hadronic matrix elements {\cite{{DGG},{Isgur1},{em}}}.
Besides providing a viable explanation for some of the difficult cases of
photohelicity amplitudes \cite{photo}, 
it is well known that one gluon generated 
configuration mixing is also able to provide viable explanation for
neutron form factor {\cite{{em},{yaouanc}}}, which cannot be accomodated
without configuration mixing in CQM.
Therefore, it becomes interesting to examine,
within the $\chi$QM, the implications
of one gluon mediated configuration mixing for flavor and spin
structure of nucleon.
In particular, we would like to examine the nucleon spin polarizations 
and various hyperon $\beta$-decay parameters,
violation of Gottfried sum rule, strange
quark content in the nucleon, fractions of quark flavor etc.
in the $\chi$QM with configuration mixing ($\chi$QM$_{gcm}$), with
and without symmetry breaking.
Further, it would be interesting to examine whether a unified
fit could be effected for spin polarization functions as well as
quark ditribution functions or not.

For the sake of readability as well to facilitate the discussion, we detail
the essentails of $\chi$QM$_{gcm}$ discussed earlier by Harleen and Gupta 
\cite{hd}.  
The basic process, in the $\chi$QM, is the
emission of a Goldstone Boson (GB) which further splits into $q \bar q$
pair, for example,

\be
  q_{\pm} \rightarrow GB^{0}
  + q^{'}_{\mp} \rightarrow  (q \bar q^{'})
  +q_{\mp}^{'}.
\ee
The effective Lagrangian describing interaction between quarks
and the octet GB and singlet $\eta^{'}$ is

\be
{\cal L} = g_8 \bar q \phi q,
\ee
where $g_8$ is the coupling constant,
\[ q =\left( \ba{c} u \\ d \\ s \ea \right)\]
and
\[ \phi = \left( \ba{ccc} \frac{\pi^o}{\sqrt 2}
+\beta\frac{\eta}{\sqrt 6}+\zeta\frac{\eta^{'}}{\sqrt 3} & \pi^+
  & \alpha K^+   \\
\pi^- & -\frac{\pi^o}{\sqrt 2} +\beta \frac{\eta}{\sqrt 6}
+\zeta\frac{\eta^{'}}{\sqrt 3}  &  \alpha K^o  \\
 \alpha K^-  &  \alpha \bar{K}^o  &  -\beta \frac{2\eta}{\sqrt 6}
 +\zeta\frac{\eta^{'}}{\sqrt 3} \ea \right). \]

SU(3) symmetry breaking is introduced by considering
different quark masses $m_s > m_{u,d}$ as well as by considering
the masses of Goldstone Bosons to be non-degenerate
 $(M_{K,\eta} > M_{\pi})$ {\cite{{song},{cheng1},{johan}}}, whereas 
  the axial U(1) breaking is introduced by $M_{\eta^{'}} > M_{K,\eta}$
{\cite{{cheng},{song},{cheng1},{johan}}}.
The parameter $a(=|g_8|^2$) denotes the transition probability
of chiral fluctuation
of the splittings  $u(d) \rightarrow d(u) + \pi^{+(-)}$, whereas 
$\alpha^2 a$ denotes the probability of transition of
$u(d) \rightarrow s  + K^{-(0)}$.
Similarly $\beta^2 a$ and $\zeta^2 a$ denote the probability of
$u(d,s) \rightarrow u(d,s) + \eta$ and
$u(d,s) \rightarrow u(d,s) + \eta^{'}$ respectively.

The one gluon exchange forces {\cite{DGG}}
generate the mixing of the octet in $(56,0^+)_{N=0}$ with
the corresponding octets in $(56,0^+)_{N=2}$,
$(70,0^+)_{N=2}$ and  $(70,2^+)_{N=2}$
harmonic oscillator bands {\cite{Isgur1}}. The
corresponding wave function of the nucleon is given by
 
 \[|B>=(|56,0^+>_{N=0} cos \theta +|56,0^+>_{N=2} sin \theta)
 cos \phi \] 
\be
 +(|70,0^+>_{N=2} cos \theta +|70,2^+>_{N=2}  sin \theta)
 sin \phi.
 \ee
In the above equation it should be noted that
$(56,0^+)_{N=2}$ does not affect the
spin-isospin structure of  $(56,0^+)_{N=0}$, therefore the mixed
nucleon wave function can be expressed in terms of  $(56,0^+)_{N=0}$
and  $(70,0^+)_{N=2}$, which we term as non trivial mixing
{\cite{mgupta1}} and is given as

\begin{equation}
\left|8,{\frac{1}{2}}^+ \right> = cos \phi |56,0^+>_{N=0}
+ sin \phi|70,0^+>_{N=2},
\end{equation}
 where
 \be
 |56,0^+>_{N=0,2} = \frac{1}{\sqrt 2}(\chi^{'} \phi^{'} +
\chi^{''} \phi^{''}) \psi^{s},
\ee

\be
|70,0^+>_{N=2} =  \frac{1}{2}[(\psi^{''} \chi^{'} +\psi^{'}
\chi^{''})\phi^{'} + (\psi^{'} \chi^{'} -\psi^{''} \chi^{''})
\phi^{''}].
\ee
 The spin and isospin wave functions, $\chi$ and $\phi$, are
given below
    \[\chi^{'} =  \frac{1}{\sqrt 2}(\uparrow \downarrow \uparrow
    -\downarrow \uparrow \uparrow),~~~  \chi^{''}
    =  \frac{1}{\sqrt 6} (2\uparrow \uparrow \downarrow
  -\uparrow \downarrow \uparrow
  -\downarrow \uparrow \uparrow), \] \\
\[\phi^{'}_p = \frac{1}{\sqrt 2}(udu-duu),~~~
\phi^{''}_p = \frac{1}{\sqrt 6}(2uud-udu-duu),\]
\[\phi^{'}_n = \frac{1}{\sqrt 2}(udd-dud),~~~
 \phi^{''}_n = \frac{1}{\sqrt 6}(udd+dud-2ddu).\]
 For the definition of the spatial part of the wave function,
 ($\psi^{s}, \psi^{'}, \psi^{''})$ as well as the
 definitions of the spatial overlap integrals, we  refer the
 reader to references {\cite{{yaouanc},{mgupta1}}.

 The contribution to the proton spin in the $\chi$QM$_{gcm}$, using the
wavefunctions defined in Equations (4)-(6), can be written as

\be
   \Delta u ={cos}^2 \phi \left[\frac{4}{3}-\frac{a}{3}
   (7+4 \alpha^2+ \frac{4}{3} \beta^2
   + \frac{8}{3} \zeta^2)\right]
   + {sin}^2 \phi \left[\frac{2}{3}-\frac{a}{3} (5+2 \alpha^2+
  \frac{2}{3} \beta^2 + \frac{4}{3} \zeta^2)\right], 
\ee

\be
  \Delta d ={cos}^2 \phi \left[-\frac{1}{3}-\frac{a}{3} (2-\alpha^2-
  \frac{1}{3}\beta^2- \frac{2}{3} \zeta^2)\right]  + {sin}^2 \phi
  \left[\frac{1}{3}-\frac{a}{3} (4+\alpha^2+
  \frac{1}{3} \beta^2 + \frac{2}{3} \zeta^2)\right], 
\ee
and

\be
   \Delta s = -a \alpha^2,
\ee   
the essential details of the derivation are presented in the Appendix A.      
The  SU(3) symmetric calculations can easily be obtained from
Equations (7), (8), (9) by considering $\alpha, \beta = 1$. The
corresponding equations can be expressed as

 \be
   \Delta u ={cos}^2 \phi \left[\frac{4}{3}
   -\frac{a}{9} (37+8 \zeta^2)\right]  + {sin}^2 \phi
  \left[\frac{2}{3}-\frac{a}{9} (23+4 \zeta^2)\right], 
\ee

\be
  \Delta d ={cos}^2 \phi \left[-\frac{1}{3}
  -\frac{2a}{9} (\zeta^2-1)\right]  + {sin}^2 \phi
  \left[\frac{1}{3}-\frac{a}{9} (16+2 \zeta^2)\right], 
\ee
and

\be
   \Delta s = -a.
\ee
After having examined the effect of one gluon exchange
inspired configuration mixing on the spin polarizations of various quarks
$\Delta u,~ \Delta d$ and $\Delta s$, we can calculate the
following quantities

\be
G_A/G_V = \Delta_3=\Delta u-\Delta d,
\ee

\be
\Delta_8=\Delta u+\Delta d-2 \Delta s.
\ee

Similarly the hyperon $\beta$ decay parameters
{\cite{{PDG},{close},{hyp1},{hyp2}}} can
also be expressed in terms of the spin polarization functions, for example,

\be
\Delta_3=\Delta u-\Delta d=F+D,
\ee

\be
\Delta_8=\Delta u+\Delta d-2 \Delta s=3F-D.
\ee

Before we present our results it is perhaps desirable to discuss
certain aspects of the symmetry breaking parameters employed here.
As has been considered by Cheng and Li {\cite{cheng}}, the singlet
octet symmetry breaking parameter
$\zeta$ is related to $\bar u- \bar d$ asymmetry
{\cite{{NMC},{E866},{GSR1}}, we have also
taken $\zeta$ to be responsible for the $\bar u-\bar d$ asymmetry
in the $\chi$QM with SU(3) symmetry breaking and configuration
mixing.
Further the parameter $\zeta$ is constrained 
\cite{{NMC},{E866},{johan}}
by the expressions $\zeta=-0.7-\beta/2$ and $\zeta=-\beta/2$
for the NMC and E866 experiments respectively, which essentially
represent the fitting of deviation from Gottfried sum rule {\cite{GSR1}}.

In Table 1, we have presented the results of our calculations
pertaining to spin polarization functions
$\Delta u, ~\Delta d, ~\Delta s$ and related parameters whereas, in
Table 2, the corresponding hyperon $\beta$-decay parameters
dependent on spin polarizations functions have been presented.
First of all we have carried out a  $\chi^2$ fit for $\chi$QM with SU(3) 
symmetry breaking and configuration mixing for the spin polarization
functions $\Delta u, ~\Delta d, ~\Delta s, ~G_A/G_V$ and other
related parameters as well as quark distribution functions. The value of 
the mixing angle is taken to be
$\phi \simeq$ 20$^o$, a value dictated by consideration of
neutron charge radius {\cite{{em},{yaouanc}}}.
In the table, however,  we have considered a few
more values of the mixing parameter $\phi$ in order to study the variation 
of spin distribution functions with  $\phi$.
The parameter $a$ is taken to be 0.1, as considered by other 
authors \cite{{cheng},{eichten},{song},{johan}}. 
The  symmetry breaking parameters obtained from  $\chi^2$ fit are 
$\alpha=.4$ and $\beta=.7$,
 both for the data corresponding to most recent E866 {\cite{E866}} as well as  
NMC {\cite{NMC}}. 
Further, while presenting the results of  SU(3) symmetry 
breaking case without configuration mixing $(\phi=0^o)$, 
we have used the same values 
of parameters  $\alpha$ and  $\beta$, primarily to understand the role of
configuration mixing for this case. 
The SU(3) symmetry calculations based on Equations (10), (11) and (12)
are obtained by taking $\alpha= \beta=1, \phi=20^o$ and  
$\alpha= \beta=1, \phi=0^o$ respectively for with and without 
configuration mixing. 
For the sake of completion, we have also presented the results of
CQM with and without configuration mixing. The spin
polarization functions of which 
can  easily be found from equations (7), (8) and (9), for example,

\be
 \Delta u = cos^2 \phi[\frac{4}{3}] + sin^2 \phi[\frac{2}{3}],
\ee

\be
 \Delta d = cos^2 \phi[-\frac{1}{3}] + sin^2 \phi[\frac{1}{3}],
\ee

and

\be
 \Delta s = 0.
\ee

A general look at Table 1,  makes it clear that we have
been able to get an excellent fit to the spin polarization
data for the values of symmetry
breaking parameters $\alpha=.4,~ \beta=.7$ obtained by
$\chi^2$ minimization.
It is perhaps desirable to mention that the spin distribution functions
$\Delta u, ~\Delta d, ~\Delta s$ show much better agreement with data
when the contribution of anomaly {\cite{anomaly}} is included.
Similarly in Table 2, we find that the success of the fit obtained with
$\alpha=.4,~ \beta=.7$ hardly leaves anything to desire. The agreement is
striking in the case of parameters F and D. We therefore conclude
that the $\chi$QM with SU(3) and axial  U(1) symmetry breakings along with
configuration mixing generated by one gluon exchange forces provides
a satisfactory description of the spin polarization functions and the
hyperon $\beta$ decay parameters.

 In order to appreciate the role of
configuration mixing in affecting the fit, we first  compare the
results of CQM with those of CQM$_{gcm}$ {\cite{hd}}.
One observes that
configuration mixing corrects the result of the quantities
in the right direction but this is not to the desirable level.
Further, in order to understand the role of configuration mixing
and SU(3) symmetry with and without breaking  in $\chi$QM,
we can compare the results of $\chi$QM with SU(3)
symmetry to those of $\chi$QM$_{gcm}$ with SU(3)
symmetry. Curiously $\chi$QM$_{gcm}$ compares unfavourably with  $\chi$QM
in case of most of the calculated quantities.
This indicates that configuration mixing alone is not enough to generate
 an appropriate fit in  $\chi$QM.
However when  $\chi$QM$_{gcm}$ is used with SU(3) and axial U(1) symmetry 
breakings
then the results show uniform improvement over the corresponding
results of $\chi$QM with  SU(3) and axial U(1) symmetry breakings.
 In particular,
the agreement with data, in the case of
$G_A/G_V, ~\Delta_8$, F, D, $f_s$ and $f_3/f_8$, is quite striking.
To summarize the  discussion of these results,
one finds that both configuration mixing
and symmetry breaking are very much needed to fit the data within
$\chi$QM.

In view   of the fact that flavor structure of nucleon is not affected
by configuration mixing, it would seem that the results of  $\chi$QM
with SU(3) breaking will be exactly similar to those of  $\chi$QM$_{gcm}$
with SU(3) breaking. However, as mentioned earlier, one of the purpose of
the present communication is to have a unified fit to spin polarization
functions as well as quark distribution functions, therefore we have 
calculated the quark distribution functions with the symmetry breaking
parameters obtained by $\chi^2$ fit in the case of $\chi$QM with symmetry 
breaking and configuration mixing both for NMC and E866 data.
To that end, we first mention the quantities which we
have calculated. The basic quantities of interest in this case are
the unpolarized quark distribution functions,
particularly the antiquark contents given as under {\cite{johan}}

\be
\bar u =\frac{1}{12}[(2 \zeta+\beta+1)^2 +20] a,
\ee

\be
\bar d =\frac{1}{12}[(2 \zeta+ \beta -1)^2 +32] a,
\ee

\be
\bar s =\frac{1}{3}[(\zeta -\beta)^2 +9 {\alpha}^{2}] a.
\ee
The quark numbers in the proton are presented as
\be
u-\bar u=2, ~~~d-\bar d=1, ~~~ s-\bar s=0,
\ee
where the quark and the antiquark numbers of a given flavor, in the quark sea,
are equal.

There are important experimentally
measurable quantities dependent on the above distributions.
The deviation from the Gottfried sum rule {\cite{GSR1}} is one such
quantity which measures the 
 asymmetry between the $\bar u$ and $\bar d$ quarks in
the nucleon sea. In the $\chi$QM the deviation of
Gottfried sum rule from 1/3rd is expressed as

\be
\left [\int_{0}^{1} dx \frac{F_2^p(x)-F_2^n(x)}{x} -\frac{1}{3}
\right ] =\frac{2}{3} (\bar u-\bar d).
\ee
Similarly the $\bar u/\bar d$, which can be measured through the
ratio of muon pair production cross sections $\sigma_{pp}$  
and $\sigma_{pn}$, is also an important   parameter which gives
an insight into the $\bar u, ~\bar d$ content {\cite{baldit}}.
The other quantities of interest is the quark flavor fraction in a 
proton, $f_q$,  defined as

\be
f_q=\frac{q+\bar q}{[\sum_{q} (q+\bar q)]},
\ee
where q's stand for the quark numbers in the proton.
Also we have calculated the ratio of the total strange sea to
the light antiquark contents given by

\be
\frac{2 \bar s}{\bar u+ \bar d},
\ee
and the ratio of the total strange sea to the light quark
contents given by

\be
\frac{2 \bar s}{u+d}.
\ee

The above mentioned quantities based on quark distribution functions
have been calculated using the set of parameters,
$\alpha=.4$ and $\beta=.7$, which minimizes the $\chi^2$ fit for the spin
distribution functions and quark distribution functions in the $\chi$QM with
SU(3) symmetry breaking as well as configuration mixing .
The results of our calculations are presented in Table 3.
The general survey of Table 3 immediately makes it clear that the 
success achieved in the case of spin polarization functions is
very well maintained in this case also. Apparently it would seem
that $\chi$QM$_{gcm}$ with SU(3) symmetry breaking would not
add anything to the success in
$\chi$QM with SU(3) symmetry breaking. However, as has been mentioned
earlier also that one of the purpose of the present communication is to have
a unified fit to spin and quark distribution functions so we have presented 
the results of our calculations with same symmetry breaking parameters 
for quark distribution functions.
The calculated values hardly leave anything to be desired  both for the 
NMC and E866 data. 
Our results show considerable
improvement in the case of ratio of the total strange sea
to the light antiquark contents ($\frac{2 \bar s}{\bar u+\bar d}$)
whereas there is a big improvement in the case of 
 ratio of the total strange sea to the light  quark contents
($\frac{2 \bar s}{u+d}$), the strange flavor fraction ($f_s$) and $f_3/f_8$ 
in comparison
to $\chi$QM with SU(3) symmetry. 

It is perhaps desirable to compare our results with those of Cheng and Li,
derived by considering SU(3) and axial U(1) breakings \cite{cheng1}. 
Before we compare our results, it needs to be mentioned
that Cheng and Li have considered the  NMC data only. It may be
 mentioned that in the present calculations, for the NMC data, 
the parameter $\zeta$ is constrained by the 
relation $\zeta=-0.7-\beta/2$, following from the fitting of $\bar u/\bar d$, 
considered by other authors as well \cite{johan},  whereas Cheng and Li 
do not put any restriction
on $\zeta$. Therefore, from Table 4, it can be seen that except for 
$\bar u/\bar d$, there is a broad agreement
between the two models. However, one must keep in mind that the 
$\chi$QM$_{gcm}$ is able to give a good fit to the E866 data as well.

To summarize, we have investigated the implications of configuration mixing
and SU(3) symmetry breaking, for proton spin and flavor structure. We
find that $\chi$QM$_{gcm}$ with SU(3) symmetry breaking is able to give a 
satisfactory unified fit for spin and quark distribution functions, with
the symmetry breaking parameters $\alpha=.4$, $\beta=.7$ and the mixing angle
$\phi=20^o$,  both for NMC as well as the most recent E866 data.
In particular, the agreement in the case of  
$G_A/G_V, ~\Delta_8$, F, D, $f_s$ and $f_3/f_8$,  is quite striking. 
 For a better appreciation of the role of configuration mixing,
we have also carried out corresponding calculations in the case of CQM with
configuration mixing and also in $\chi$QM with SU(3) symmetry and 
SU(3) symmetry breaking 
without configuration mixing, the latter being carried out with the same 
values of the symmetry breaking parameters, $\alpha=.4$ and $\beta=.7$.
It is found that configuration mixing improves the CQM results, however in
the case of  $\chi$QM with SU(3) symmetry the results become worse. The 
situation changes completely when SU(3) symmetry breaking and 
configuration mixing are included simultaneously. 
Thus, it seems that both configuration mixing as well as symmetry 
breaking are very much needed to fit the data within $\chi$QM.

\vskip .2cm
  {\bf ACKNOWLEDGMENTS}\\
H.D. would like to thank CSIR, Govt. of India, for
 financial support and the chairman,
 Department of Physics, for providing facilities to work
 in the department.

\newpage

\begin{center}
{\bf APPENDIX} 
\end{center}

The spin structure of a nucleon, following Linde {\it et al.} \cite{johan}, 
is defined as \linebreak $\hat B \equiv <B|N|B>$,
where $|B>$ is the nucleon wavefunction and $N$ is the number operator
given by

\[ N=n_{u^{\uparrow}}u^{\uparrow} + n_{u^{\downarrow}}u^{\downarrow} +
n_{d^{\uparrow}}d^{\uparrow} + n_{d^{\downarrow}}d^{\downarrow} +
n_{s^{\uparrow}}s^{\uparrow} + n_{s^{\downarrow}}s^{\downarrow}, \]
where the coefficients of the $q^{\uparrow \downarrow}$ are the number of 
$q^{\uparrow \downarrow}$ quarks. 
The spin structure of the `mixed' nucleon, defined through the
 Equation (4), is given by
\[ \left< 8,{\frac{1}{2}}^+|N|8,{\frac{1}{2}}^+\right>=cos^2 \phi
<56,0^+|N|56,0^+>+sin^2 \phi<70,0^+|N|70,0^+>,~~(A1) \]

\[ <56,0^+|N|56,0^+>=\frac{5}{3} u^{\uparrow} +\frac{1}{3} u^{\downarrow}+
\frac{1}{3} d^{\uparrow} +\frac{2}{3} d^{\downarrow}, ~~~~~~~~~~~~(A2)\]
 and
\[ <70,0^+|N|70,0^+>=\frac{4}{3} u^{\uparrow} +\frac{2}{3} u^{\downarrow}+
\frac{2}{3} d^{\uparrow} +\frac{1}{3} d^{\downarrow}, ~~~~~~~~~~~~(A3)\]
where we have used Equations (5) and (6) of the text.

In the  $\chi$QM, the basic process is the emission of a Goldstone Boson
 which further splits into $q \bar q$ pair as mentioned in Equation (1) of the
text. Following Linde {\it et al.} the spin structure
after one interaction can be obtained by  substituting for every quark, for
example,

\[ q^{\uparrow} \rightarrow P_q q^{\uparrow} + |\psi(q^{\uparrow})|^2, \]
where $P_q$ is the probability of no emission of GB from a q quark and the 
probabilities of transforming a $q^{\uparrow \downarrow}$ quark are
$|\psi(q^{\uparrow})|^2$, given as

\[ |\psi(u^{\uparrow})|^2=\frac{a}{6}(3+\beta^2+2 \zeta^2)u^{\downarrow}+
a d^{\downarrow}+a \alpha^2 s^{\downarrow},     ~~~~~~~~~~~~(A4) \]

\[ |\psi(d^{\uparrow})|^2=a u^{\downarrow}+ 
\frac{a}{6}(3+\beta^2+2 \zeta^2)d^{\downarrow}+ a \alpha^2 s^{\downarrow}, 
                                              ~~~~~~~~~~~~(A5) \]

\[ |\psi(s^{\uparrow})|^2=   a \alpha^2 u^{\downarrow}+ 
a \alpha^2 d^{\downarrow}+\frac{a}{3}(2 \beta^2+\zeta^2)s^{\downarrow}.
 ~~~~~~~~~~~~~~~(A6) \]
For the definitions of $\alpha, \beta$ and $\zeta$ we refer the 
readers to the text.

Using Equations (A1)-(A6), we obtain

\[ \hat B=cos^2 \phi \left [ \frac{5}{3}(P_u u^{\uparrow} + 
|\psi(u^{\uparrow})|^2)+
\frac{1}{3}(P_u u^{\downarrow} + |\psi(u^{\downarrow})|^2)+
\frac{1}{3}(P_d d^{\uparrow} + |\psi(d^{\uparrow})|^2)+ 
 \frac{2}{3}(P_d d^{\downarrow} + |\psi(d^{\downarrow})|^2) \right ] \]
\[+sin^2 \phi \left [ \frac{4}{3}(P_u u^{\uparrow} + |\psi(u^{\uparrow})|^2)+
\frac{2}{3}(P_u u^{\downarrow} + |\psi(u^{\downarrow})|^2)+
\frac{2}{3}(P_d d^{\uparrow} + |\psi(d^{\uparrow})|^2)+
\frac{1}{3}(P_d d^{\downarrow} + |\psi(d^{\downarrow})|^2) \right ].~~(A7)\]

The spin polarization for any quark is defined as $ \Delta q=n_{q^{\uparrow}}-n_{q^{\downarrow}}+n_{\bar q^{\uparrow}}-n_{q^{\downarrow}}.$
Using the spin structure from Equation (A7) 
we can calculate the spin polarizations, which come out to be

\[   \Delta u ={cos}^2 \phi \left[\frac{4}{3}-\frac{a}{3}
   (7+4 \alpha^2+ \frac{4}{3} \beta^2
   + \frac{8}{3} \zeta^2)\right]
   + {sin}^2 \phi \left[\frac{2}{3}-\frac{a}{3} (5+2 \alpha^2+
  \frac{2}{3} \beta^2 + \frac{4}{3} \zeta^2)\right], ~~~~~~(A8)\]

\[  \Delta d ={cos}^2 \phi \left[-\frac{1}{3}-\frac{a}{3} (2-\alpha^2-
  \frac{1}{3}\beta^2- \frac{2}{3} \zeta^2)\right]  + {sin}^2 \phi
  \left[\frac{1}{3}-\frac{a}{3} (4+\alpha^2+
  \frac{1}{3} \beta^2 + \frac{2}{3} \zeta^2)\right], ~~~~~~~(A9)\]
and

\[   \Delta s = cos^2 \phi[-a \alpha^2] + sin^2 \phi[-a \alpha^2]= 
-a \alpha^2.~~~~~~~~~~~~~~~~~~~~~~~~~~~(A10)\]

It is interesting to note that, in the case of $\Delta s$, both 
$|56>$ and $|70>$ 
wavefunctions are contributing in the same manner. It can be easily understood
when one considers Equations (A4) and (A5). As is evident, in these equations
both $u^{\uparrow}$ and $d^{\uparrow}$ contribute to $s^{\downarrow}$ in the 
same manner, similarly    $u^{\downarrow}$ and  $d^{\downarrow}$ contribute
to  $s^{\uparrow}$ in the same manner. When this is used along with Equations
(A2) and (A3), one can immediately find out that $\Delta s$ comes 
out to be $-a \alpha^2$.

It is interesting to compare the spin polarizations given by
 Equations  (A8) and (A9) to similar ones derived by  Linde {\it et al.}
\cite{johan} (Equations B1 and B2) in the case of wavefunction with 
quark-gluon mixing(Equations 17 and 18 of reference \cite{johan}). 
Since the $|56>$ part of the wavefunction is same in both the cases, we 
compare the  term corresponding to the $sin^2 \phi$ coefficients.
In  Equations (A8) and (A9),
in comparision to the coefficients of $sin^2 \phi$, 
the corresponding term in the case of  Linde {\it et al.} has
proportionality factor of -1/3 despite a different corresponding 
spin structure. 
This can  be easily understood when one recognises
that the contributions to $\Delta u$ and $\Delta d$ are 
proportional to $[\frac{4}{3}u^{\uparrow}-\frac{2}{3}u^{\downarrow}]$ and 
 $[\frac{2}{3}d^{\uparrow}-\frac{1}{3}d^{\downarrow}]$ for the $|70>$
wavefunction. Interestingly the spin polarizations for the 
gluonic wavefunction considered by Linde {\it et al.}, 
are proportional to  
 $[\frac{8}{9}u^{\uparrow}-\frac{10}{9}u^{\downarrow}]$ and
 $[\frac{4}{9}d^{\uparrow}-\frac{5}{9}d^{\downarrow}]$, which has a 
proportionality factor of -1/3 as compared to our $|70>$ wavefunction. Thus, 
the contributions of the spin polarizations corresponding to the 
coefficients of $\sin^2 \phi$ are proportional.

\pagebreak

\begin{table}
{\tiny
\begin{center}
\begin{tabular}{|c|c|c|c|c|c|c|c|c|c|c|c|c|}       \hline
 & & \multicolumn{5}{c|} {Without configuration mixing} &
\multicolumn{6}{c|} {With configuration mixing}\\ \cline{3-13} 
Para- & Expt & CQM & \multicolumn{2}{c|} {$\chi$QM}  &
\multicolumn{2}{c|} {$\chi$QM}
& $\phi$ & CQM$_{gcm}$ & \multicolumn{2}{c|} {$\chi$QM$_{gcm}$}     &
\multicolumn{2}{c|} {$\chi$QM$_{gcm}$} \\
meter & value&   & \multicolumn{2}{c|} {with SU(3)}  &
\multicolumn{2}{c|} {with SU(3)} & & & \multicolumn{2}{c|} {with SU(3)}  &
\multicolumn{2}{c|} {with SU(3)} \\
 & &   & \multicolumn{2}{c|} {symmetry}  &
\multicolumn{2}{c|} {symmetry} & &
& \multicolumn{2}{c|} {symmetry}  &
\multicolumn{2}{c|} {symmetry} \\  
&&&\multicolumn{2}{c|}{}& \multicolumn{2}{c|} {breaking} &&&
\multicolumn{2}{c|}{}& \multicolumn{2}{c|} {breaking} \\ \hline
& & & NMC  & E866  & NMC &E866 & & & NMC & E866 & NMC & E866 \\
\cline{4-7} \cline{10-13}
 & & & & & & & 20$^o$ & 1.26 &
 .74 & .76 & .90, .86$^*$ & .92, .88$^*$  \\
 $\Delta$ u & 0.85 $\pm$ 0.05  & 1.33 & .79 & .81 & .96 & .99 & 18$^o$ 
& 1.27 &
.75 &.77 & .91, .87$^*$ & .93, .89$^*$  \\
& {\cite{adams}} & & & & & & 16$^o$ & 1.28 &
.76 & .78 & .92, .88$^*$ & .94, .90$^*$ \\
 & & & & & & & 14$^o$ & 1.29 &
.77 & .79 & .93, .89$^*$ & .95, .91$^*$ \\ \hline

 & & & & & & & 20$^o$ &
 -0.26 & -0.30 & -0.31  & -0.32, -0.36$^*$ & -0.34, -0.38$^*$ \\
 $\Delta$ d & -0.41  $\pm$ 0.05  & -0.33 & -0.35 & -0.37 & -0.40 & -0.41 & 
18$^o$ & -0.27 &-0.31 & -0.32 & -0.33, -0.37$^*$ & -0.35, -0.39$^*$ \\
   & {\cite{adams}} & & & & & & 16$^o$ &
  -0.28 & -0.32 &-0.33 & -0.34, -0.38$^*$ & -0.36. -0.40$^*$ \\
   & & & & & & & 14$^o$ &
 -0.29 & -0.33 & -0.34 & -0.35, -0.39$^*$ & -0.37, -0.41$^*$ \\ \hline

&&&&&&&&&&&& \\
$\Delta$ s &-0.07  $\pm$ 0.05 & 0 & -0.1 & -0.12 & -0.02 & -0.02 & &
0 & -0.1 & -0.12 & -0.02, -0.06$^*$ &-0.02, -0.06$^*$ \\  
& {\cite{adams}} &&&&&&&&&&& \\ \hline

 & & & & & & & 20$^o$ &
1.52 & 1.04 & 1.07 & 1.22, 1.22$^*$ & 1.26, 1.26$^*$ \\
$G_A/G_V$ & 1.267  $\pm$ .0035 & 1.66 & 1.14 & 1.18 & 1.35 & 1.40 & 18$^o$ &
1.54 & 1.06 & 1.09 & 1.24, 1.24$^*$ & 1.28, 1.28$^*$  \\
 & {\cite{PDG}}& & & & & & 16$^o$ &         
1.56 & 1.08 & 1.11 & 1.26, 1.26$^*$ & 1.30, 1.30$^*$  \\
 & & & & & & & 14$^o$ &
1.58 & 1.10 & 1.13 & 1.28, 1.28$^*$ & 1.32, 1.32$^*$ \\   \hline

 & & & & & & & 20$^o$ &
1 & .64 & .69 & .62 & .62 \\
$\Delta_8$ & .58  $\pm$ .025 & 1 & .64 & .68 & .60 & .62  & 18$^o$ &
1 & .64 & .69 & .62 & .62  \\
 & {\cite{PDG1}}& & & & & & 16$^o$ &         
1 & .64 & .69 & .62 & .62  \\
 & & & & & & & 14$^o$ &
1 & .64 & .69 & .62 & .62 \\   \hline

\end{tabular}
\end{center}}
* {\small Values after inclusion of the contribution from anomaly
{\cite{anomaly}}.}
\caption{The calculated values of
spin polarization functions $\Delta u, ~\Delta d, ~\Delta s$,
and quantities dependent  on these: $G_A/G_V$ and $\Delta_8$ 
both for  NMC and E866 data with the symmetry breaking 
parameters obtained by $\chi^2$ minimization in the $\chi$QM with 
 one gluon generated configuration mixing ($\chi$QM$_{gcm}$) and
SU(3) symmetry breaking.}
\end{table}

\begin{table}
{\tiny
\begin{center}
\begin{tabular}{|c|c|c|c|c|c|c|c|c|c|c|c|c|}       \hline
 & & \multicolumn{5}{c|} {Without configuration mixing} &
\multicolumn{6}{c|} {With configuration mixing}\\ \cline{3-13} 
Para- & Expt & CQM & \multicolumn{2}{c|} {$\chi$QM}  &
\multicolumn{2}{c|} {$\chi$QM}
& $\phi$ & CQM$_{gcm}$ & \multicolumn{2}{c|} {$\chi$QM$_{gcm}$}     &
\multicolumn{2}{c|} {$\chi$QM$_{gcm}$} \\
 meter & value&   & \multicolumn{2}{c|} {with SU(3)}  &
\multicolumn{2}{c|} {with SU(3)} & & & \multicolumn{2}{c|} {with SU(3)}  &
\multicolumn{2}{c|} {with SU(3)} \\
 & &   & \multicolumn{2}{c|} {symmetry}  &
\multicolumn{2}{c|} {symmetry} & &
& \multicolumn{2}{c|} {symmetry}  &
\multicolumn{2}{c|} {symmetry} \\  
&&&\multicolumn{2}{c|}{}& \multicolumn{2}{c|} {breaking} &&&
\multicolumn{2}{c|}{}& \multicolumn{2}{c|} {breaking} \\ \hline
& & & NMC  & E866  & NMC &E866 & & & NMC & E866 & NMC & E866 \\
\cline{4-7} \cline{10-13}

 & & & & & & & 20$^o$ & 1.52 &
 1.04 & 1.07 & 1.22 & 1.26  \\
 F+D & 1.26   & 1.66 & 1.14 & 1.18 & 1.36 & 1.40 & 18$^o$ 
& 1.54 &
1.06 & 1.09 & 1.24 & 1.28  \\
& & & & & & & 16$^o$ & 1.56 &
1.08 & 1.11 & 1.26 & 1.30 \\
 & & & & & & & 14$^o$ & 1.58 &
1.10 & 1.13 & 1.28 & 1.32 \\ \hline

& & & & & & & 20$^o$ & .93 &
 .63 & .65 & .71 & .73  \\
 F+D/3 & .718   & 1 & .677 & .703 & .78 & .80 & 18$^o$ 
& .94 &
.636 &.66 & .72 & .74  \\
& & & & & & & 16$^o$ & .95 &
.646 & .67 & .73 & .75 \\
 & & & & & & & 14$^o$ & .96 &
.656 & .68 & .74 & .76 \\ \hline

& & & & & & & 20$^o$ & -.26 &
 -.20 & -.19 & -.30 & -.32  \\
 F-D & -.34   & -.33 & -.25 & -.25 & -.38 & -.39 & 18$^o$ 
& -.27 &
-.21 & -.20 & -.31 & -.33  \\
& & & & & & & 16$^o$ & -.28 &
-.22 & -.21 & -.32 & -.34 \\
 & & & & & & & 14$^o$ & -.29 &
-.23 & -.22 & -.33 & -.35 \\ \hline

& & & & & & & 20$^o$ & .33  &
 .21 & .23 & .21 & .21  \\
 F-D/3 & .25   & .33 & .21 & .23 & .20 & .21 & 18$^o$ 
& .33  &
.21 & .23 & .21 & .21  \\
& & & & & & & 16$^o$ & .33 &
.21 & .23 & .21 & .21  \\
 & & & & & & & 14$^o$ & .33 &
 .21 & .23 & .21 & .21 \\ \hline

& & & & & & & 20$^o$ & .71 &
 .68 & .70 & .61 & .59  \\
 F/D & .575   & .67 & .64 & .65 & .56 & .56 & 18$^o$ 
& .70 &
.67 &.69 & .60 & .58  \\
& & & & & & & 16$^o$ & .69 &
.66 & .68 & .59 & .57 \\
 & & & & & & & 14$^o$ & .68 &
.65 & .67 & .58 & .56 \\ \hline

& & & & & & & 20$^o$ & .63 &
 .42 & .44 & .46 & .47  \\
 F & .462   & .665 & .445 & .465 & .49 & .505 & 18$^o$ 
& .635 &
.425 &.445 & .465 & .475  \\
& & & & & & & 16$^o$ & .64 &
.43 & .45 & .47 & .48 \\
 & & & & & & & 14$^o$ & .645 &
.435 & .455 & .475 & .485 \\ \hline

& & & & & & & 20$^o$ & .89 &
 .62 & .63 & .76 & .79  \\
 D & .794  & 1 & .695 & .715 & .87 & .895 & 18$^o$ 
& .905 &
.635 &.645 & .775 & .805  \\
& & & & & & & 16$^o$ & .920 &
.65 & .66 & .79 & .82 \\
 & & & & & & & 14$^o$ & .935 &
.665 & .675 & .805 & .835 \\ \hline

\end{tabular}
\end{center}}
\caption{The calculated values of hyperon $\beta$ decay parametres in the
$\chi$QM
with and without configuration mixing as well as with and without SU(3) 
symmetry breaking for the values of $\alpha$ and $\beta$ obtained by $\chi^2$ 
minimization in the case of $\chi$QM$_{gcm}$ with
SU(3) symmetry breaking for NMC and E866 data.}
\end{table}

\begin{table}
{\small
\begin{center}
\begin{tabular}{|c|c|c|c|c|c|c|}       \hline
              
Parameter & Expt & CQM & \multicolumn{2}{c|} {$\chi$QM}  &
\multicolumn{2}{c|} {$\chi$QM} \\
 & value&   & \multicolumn{2}{c|} {with SU(3)}  &
\multicolumn{2}{c|} {with SU(3)}  \\
 & &   & \multicolumn{2}{c|} {symmetry}  &
\multicolumn{2}{c|} {symmetry}  \\  
&&&\multicolumn{2}{c|}{}& 
\multicolumn{2}{c|} {breaking} \\ \hline                            
& & & NMC  & E866  & 
NMC &E866                               \\
\cline{4-7}
 
$\bar u$ &  & & .168 & .21 &.168  &  .21  \\

$\bar d$ &  & & .315 & .33 & .315 &  .33  \\

$\bar s$ &  & & .46 & .45 & .15 &   .10  \\

$\bar d-\bar u$ &.147 $\pm$ .024 {\cite{NMC}} & 0 &  .147  &
.12 & .147 & .12  \\
                & .100 $\pm$ .015 \cite{E866} &&&&& \\

$\bar u/\bar d$ & 0.51 $\pm$ 0.09 {\cite{baldit}} & 1 & .53 &
.63 & .53 &  .63  \\
                & 0.67 $\pm$ 0.06 \cite{E866} &&&&& \\

$I_G$ & .235  $\pm$ .005 & 0.33 & .235 & .253 & .235 & .253  \\
      &.266     $\pm$ .005   &&&&& \\

$\frac{2 \bar s}{\bar u+ \bar d}$ & .477 $\pm$ .051 {\cite{ao}} &
&1.9 & 1.66  &  .62 & .38 \\

$\frac{2 \bar s}{u+d}$ & .099 $\pm$ .009 {\cite{ao}} & 0 & .26 &
 .25 &   .09 & .06  \\

$f_u$ &  & & .48 & .49 & .55 &   .56  \\

$f_d$ & & & .33 & .33 & .38 &  .39  \\

$f_s$ &  .10 $\pm$ 0.06 {\cite{ao}} & 0 & .19 &
.18 &   .07 & .05  \\

$f_3=$  & & & .15 & .15 &   .17 & .18 \\
$f_u-f_d$  &  & & && &  \\

$f_8=$ & & & .43 & .46 &    .79 & .86 \\
$f_u+f_d-2 f_s$ &   & && & &   \\

$f_3/f_8$ & .21 $\pm$ 0.05 {\cite{cheng}} & .33 & .33 & .33 &
.21 &   .21 \\
\hline

\end{tabular}
\end{center}}
\caption{The calculated values of quark distribution functions and other 
dependent quantities as calculated in the $\chi$QM with and without SU(3) 
symmetry breaking both for NMC and E866 data, with the same values of symmetry
breaking parameters as used in spin distribution functions and hyperon
$\beta$ decay parameters.}
\end{table}

\begin{table}
\begin{center}
\begin{tabular}{|c|c|c|c|c|}       \hline
              
Parameter & Expt value & Cheng and Li {\cite{cheng1}} & 
\multicolumn{2}{c|} {Present results} \\ \cline{4-5}   
&&& NMC & E866 \\ \hline

 $\Delta$ u & 0.85 $\pm$ 0.05 {\cite{adams}} & .87 & .87 & .89 \\

 $\Delta$ d & -0.41  $\pm$ 0.05  {\cite{adams}} & -0.41 & -0.37 & -0.39 \\

$\Delta$ s &-0.07  $\pm$ 0.05  {\cite{adams}} & -0.05 & -0.06 & -0.06 \\

$G_A/G_V$ & 1.267  $\pm$ .0035  {\cite{PDG}} & 1.28 & 1.24 & 1.28 \\

F/D & .575 & .57 & .60 & .58 \\

3F-D & .60 & .57 & .62 & .62 \\

$\bar d-\bar u$ &.147 $\pm$ .024 {\cite{NMC}}  &
.15 & .147 & .12  \\
                & .100 $\pm$ .015 \cite{E866} &&& \\

$\bar u/\bar d$ & 0.51 $\pm$ 0.09 {\cite{baldit}} &  .63  & .53 &  .63  \\
                & 0.67 $\pm$ 0.06 \cite{E866} &&& \\

$\frac{2 \bar s}{\bar u+ \bar d}$ & .477 $\pm$ .051 {\cite{ao}} & .60  
&  .62 & .38 \\

$f_s$ &  .10 $\pm$ 0.06 {\cite{ao}} & .09 &   .07 & .05  \\

$f_3/f_8$ & .21 $\pm$ 0.05 {\cite{cheng}} & .20 &
.21 &   .21 \\     \hline

\end{tabular}
\end{center}
\caption{Comparison of the results of $\chi$QM$_{gcm}$  for the quark spin and
flavor distribution functions at the mixing angle $\phi = 18^o$ with those of 
Cheng and Li.}
\end{table}

\end{document}